\documentclass[sigconf]{acmart}

\AtBeginDocument{%
  }

\setcopyright{acmlicensed}
\copyrightyear{2018}
\acmYear{2018}
\acmDOI{XXXXXXX.XXXXXXX}
\acmConference[Conference acronym 'XX]{Make sure to enter the correct
  conference title from your rights confirmation email}{June 03--05,
  2018}{Woodstock, NY}
\acmISBN{978-1-4503-XXXX-X/2018/06}

\usepackage{hyperref}
\begin{document}

%\title{Fine-Grained Yet Efficient: Enhancing Bidirectional Information Flow for Scalable Unified Feature Interaction}
\title{Bridging Sequential and Contextual Features with a Dual-View of Fine-grained Core-Behaviors and Global Interest-Distribution}

\author{Yi Xu}
%\authornote{Both authors contributed equally to this research and are co-first authors.}
\affiliation{
  \institution{Alibaba Group}
  \city{Beijing}\country{China}
}
\email{xy397404@alibaba-inc.com}

\author{Chaofan Fan}
\affiliation{
  \institution{Alibaba Group}
  \city{Beijing}\country{China}
}
\email{fanchaofan.fcf@alibaba-inc.com}

\author{Moyu Zhang}
\affiliation{
  \institution{Alibaba International Digital Commerce Group}
  \city{Beijing}\country{China}
}
\email{zhangmoyu.zmy@alibaba-inc.com}

\author{Jinxin Hu}
\affiliation{
  \institution{Alibaba Group}
  \city{Beijing}\country{China}
}
\email{jinxin.hjx@alibaba-inc.com}
\authornote{Corresponding author}

\author{Jiahao Wang}
\affiliation{
  \institution{Alibaba International Digital Commerce Group}
  \city{Beijing}\country{China}
}
\email{wjh177423@alibaba-inc.com}

\author{Hao Zhang
}
\affiliation{
  \institution{Alibaba International Digital Commerce Group}
  \city{Beijing}\country{China}
}
\email{zh138764@alibaba-inc.com}

\author{Shizhun Wang}
\affiliation{
  \institution{Alibaba International Digital Commerce Group}
  \city{Beijing}\country{China}
}
\email{shaoan.wsz@taobao.com}

\author{Yu Zhang}
\affiliation{
  \institution{Alibaba Group}
  \city{Beijing}\country{China}
}
\email{daoji@alibaba-inc.com}

\author{Xiaoyi Zeng}
\affiliation{
  \institution{Alibaba Group}
  \city{Beijing}\country{China}
}
\email{yuanhan@taobao.com}
% \affiliation{%

\renewcommand{\shortauthors}{Trovato et al.}

\begin{abstract}
%CTR prediction remains a core challenge in recommender systems. 
Click-through rate (CTR) prediction tasks typically estimate the probability of a user clicking on a candidate item by modeling both user behavior sequence features and the item's contextual features, where the user behavior sequence is particularly critical as it dynamically reflects real-time shifts in user interest. Traditional CTR models often aggregate this dynamic sequence into a single vector before interacting it with contextual features. This approach, however, not only leads to behavior information loss during aggregation but also severely limits the model's capacity to capture interactions between contextual features and specific user behaviors, ultimately impairing its ability to capture fine-grained behavioral details and hindering models' prediction accuracy. Conversely, a naive approach of directly interacting with each user action with contextual features is computationally expensive and introduces significant noise from behaviors irrelevant to the candidate item. This noise tends to overwhelm the valuable signals arising from interactions involving more behaviors relevant to the candidate item. Therefore, to resolve the above issue, we propose a \textbf{C}ore-Behaviors and \textbf{D}istributional-Compensation Dual-View Interaction \textbf{Net}work (CDNet), which bridges the gap between sequential and contextual feature interactions from two complementary angles: a fine-grained interaction involving the most relevant behaviors and contextual features, and a coarse-grained interaction that models the user's overall interest distribution against the contextual features. By simultaneously capturing important behavioral details without forgoing the holistic user interest, CDNet effectively models the interplay between sequential and contextual features without imposing a significant computational burden. Ultimately, extensive experiments validate the effectiveness of CDNet.
\end{abstract}

\begin{CCSXML}
<ccs2012>
   <concept>a
       <concept_id>10002951.10003317.10003331.10003271</concept_id>
       <concept_desc>Information systems~Personalization</concept_desc>
       <concept_significance>500</concept_significance>
       </concept>
 </ccs2012>
\end{CCSXML}

\ccsdesc[500]{Information systems~Personalization}

\keywords{Dual-View Interaction Network; Sequential and Contextual Features Interaction; Click-Through Rate Prediction}

\maketitle
\section{Introduction}
Click-through rate (CTR) prediction is a cornerstone of modern recommender systems, which estimates the probability of a user clicking a candidate item by modeling the interaction between the user's dynamic behavior sequence and the static contextual features\cite{intro1,Zhang2022DHENAD,gui2023hiformerheterogeneousfeatureinteractions,Directed}. While contextual features describe the item's or the user's intrinsic characteristics\cite{xDeepFM,DCNV2}, the user's dynamic behavior sequences are of paramount importance as it captures the real-time evolution of user interests\cite{Wang2019SequentialRS,DIN}. Consequently, particularly in non-search recommendation scenarios, a model's predictive accuracy hinges on its ability to effectively capture the intricate interactions between these two distinct feature types\cite{interformer,onetrans,EulerNet}.

However, despite the considerable success of deep learning in this area, prevailing feature interaction models in the CTR prediction field have largely overlooked the fundamental heterogeneity between these two feature types. A common practice is to first aggregate the user sequence into a single or multiple fixed-length user interest vector\cite{longer,DIN,BERT4Rec}. The compressed representations are then used for downstream interactions with the static contextual features, often through sophisticated mechanisms like FM\cite{FM,AFM,FFM}, DNN\cite{dcn,DCNV2}, or Attention\cite{vaswani2023attentionneed,AutoInt}. This pre-aggregation strategy creates a critical information bottleneck. It not only results in the loss of granular behavioral information, but also, more importantly, it prevents the model from capturing the fine-grained interactions between specific past behaviors and the current candidate item\cite{interformer,onetrans}. As a result, the predictive contribution of the important user behaviors is diluted, ultimately capping the model's performance.

Recently, a line of feature interaction research in the CTR field has emerged to address the above issue. The most intuitive approach involves performing a full fine-grained interaction, where every behavior in the user sequence is directly interacted with the contextual features \cite{interformer,onetrans,hstu,MTGR,OnePiece}. Although this full fine-grained approach preserves maximum behavioral detail, its computational complexity grows quadratically with sequence length, rendering it impractical for long user histories. Moreover, this approach is also highly susceptible to noise, as the signals from the few truly pertinent behaviors can be easily overwhelmed by the multitude of interactions from behaviors irrelevant to the candidate item, ultimately degrading prediction quality. Although subsequent work has attempted to mitigate the computational and noise drawbacks by employing multi-window aggregation schemes\cite{interformer,huang2026hyformerrevisitingrolessequence,Climber}. However, they represent a regression to coarse-grained modeling, where the compression operation inevitably comes at the cost of fine-grained interactions with sequential and contextual features, thereby reintroducing the information loss they were designed to prevent.

Therefore, the central challenge of feature interaction still remains: how to effectively model the interactions between user behaviors and contextual features in a way that is both computationally efficient and robust to the noise from irrelevant actions. 

To tackle the above challenge, we propose a \textbf{C}ore-Behaviors and \textbf{D}istributional-Compensation Dual-View Interaction \textbf{Net}work (CDNet), which bridges the gap between sequential and contextual feature interactions from two complementary angles: a fine-grained interaction involving the most relevant behaviors and contextual features, and a coarse-grained interaction that models the user’s overall interest distribution against the contextual features. Specifically, our framework consists of two core modules:1) Fine-grained Core-Behaviors Module. Recognizing that user sequences often contain numerous behaviors irrelevant to the candidate item, we identify a subset of core behaviors from the user's historical behaviors that exhibit high relevance to the candidate item. This allows for a fine-grained interaction between these behaviors and contextual features, enhancing predictive accuracy. Simultaneously, this process drastically reduces the number of behaviors involved in the interaction, which significantly lowers the model's computational complexity. 2) Coarse-grained Interest-Distribution Module. Simply discarding the remaining non-core behaviors would lead to a significant loss of information regarding the user's broader interests. To preserve this holistic context without reintroducing behavior noise, we model the similarity distribution of the entire historical sequence relative to the candidate item, which serves as a summary of the user's global interests and can be used for coarse-grained interaction with the contextual features. By integrating the information of two complementary views, CDNet effectively models the interplay between sequential and contextual features, capturing critical behavioral details while preserving the user's overall interest profile, all while remaining computationally efficient. 

The contributions of our paper can be summarized as follows:
\begin{itemize}
    \item To the best of our knowledge, CDNet is the first work to simultaneously model fine-grained behavior interactions while using coarse-grained interactions supplementation to bridge sequential and contextual feature interactions.
    \item We propose the core behaviors and interest-distribution angles to identify core behaviors for a fine-grained interaction with contextual features and model the global interest-distribution for a coarse-grained interaction, respectively. 
    \item Extensive offline and online experiments demonstrate the superiority of our proposed framework.
\end{itemize}
%\section{Related Works}

% \begin{figure*}[t]
% % \vspace{-0.3cm}
%     % \centering
%     % 跨双栏时 width 用 \textwidth 而不是 \linewidth
%     \includegraphics[width=\textwidth]{sects/storev2_0211.pdf}
%     \caption{Overview of the proposed STOREv2.}
%     \label{fig:method_overview}
%     % \vspace{-0.5cm}
% \end{figure*}
% %\vspace{-0.3cm}

\begin{figure}[t]
\includegraphics[width=0.75\columnwidth]{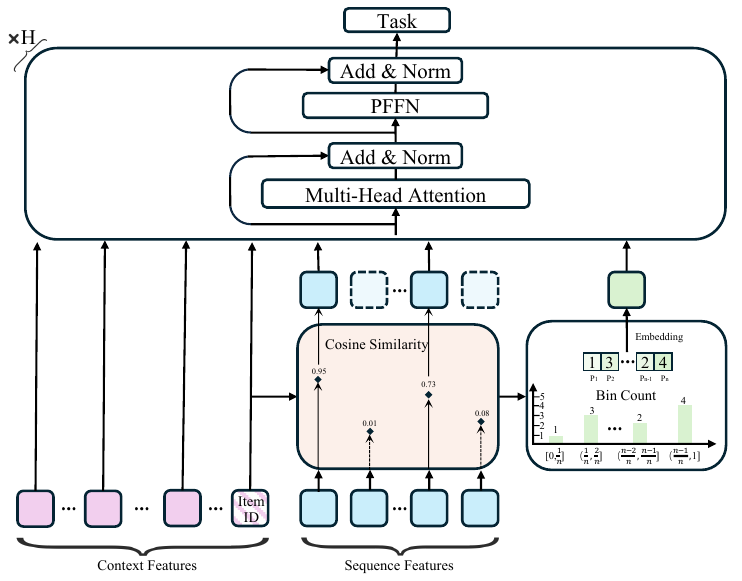}
\centering
\caption{Overview of the proposed CDNet.}
\label{fig:intro_heat}
\vspace{-0.3cm}
\end{figure}

\section{Problem Formulation}
The goal of click-through rate (CTR) prediction is to estimate the probability $\hat{y} \in [0, 1]$ that a user $u$ will click a target item $i$. This task primarily relies on modeling interactions between two main categories of features: (1) a set of relatively static and non-sequential contextual features $\mathcal{F}$, which includes user profiles (e.g., age, gender), item attributes (e.g., item ID, category ID, brand ID), and contextual information (e.g., geolocation, device); and (2) dynamic user behavior sequence features $\mathcal{S}$, which comprise the item IDs from the user's dynamic historical interactions along with their corresponding attributes (e.g., category ID and brand ID). The objective of CTR prediction can be formulated as follows:
\begin{equation}
\label{eq:task}
    \hat{y} = f(\mathcal{F}, \mathcal{S}; \theta)
\end{equation}
where $\theta$ denotes the model parameters, and $f(\cdot)$ denotes the feature interaction and prediction function.

\section{Methodology}
%\subsection{Task Defination}
As previously discussed, traditional feature interaction methods often overlook the crucial interactions between user behavior sequences and contextual features, leading to performance bottlenecks. While recent approaches attempt to capture this information through sequence unrolling or multi-window aggregation, they face a critical trade-off: the former often introduces significant computational overhead and noise, while the latter can sacrifice crucial fine-grained behavioral details, resulting in information loss. To address this challenge, this paper introduces the Core-Behaviors and Distributional-Compensation Dual-View Interaction Network (CDNet), which employs two complementary modules: a Fine-grained Core-Behaviors Module and a Global Interest-Distribution Module. The former is designed to explicitly model detailed interactions between specific user behaviors and contextual features. The latter compensates for potential information loss by modeling the user's holistic interest distribution at a coarse-grained level.

\subsection{Fine-grained Core-Behaviors Module}
To circumvent the trade-off between the high computational overhead and noise from full sequence unfolding and the information loss associated with sequence compression, we selectively identify the most core historical behaviors with respect to the candidate item. These selected behaviors are then used to perform fine-grained interactions with the contextual features, which allows for the explicit modeling of crucial interactions between core behaviors and contextual features, all while avoiding a significant increase in computational complexity.

\noindent \textbf{Differentiable Core-Behaviors Selection} Specifically, we employ the features of the target item $f_i$ as a query to distill $k$ most relevant behavioral details from the user's historical sequence $\mathcal{S} = \left\{s_1, ..., s_j, ..., s_L\right\}$, where $L$ is the length of the behavior sequence. We compute the similarity between the feature representation of the target item and that of each historical behavior. This similarity value then serves as an importance score, yielding the following score sequence: $\mathcal{A} = \left\{a_1, ..., a_j, \dots, a_L\right\}$, where $a_j=(\frac{\boldsymbol{f}_i^\top \boldsymbol{s}_j}{\lVert\boldsymbol{f}_i\rVert \lVert\boldsymbol{s}_j\rVert}+1)/2 \in [0,1]$.

Based on $\mathcal{A}$, we identify the indices of the top-$k$ most core baheviors for the item $i$ to form the core behavior set:
\begin{equation}
\mathcal{I}_{\mathbf{k}} = \text{top-k}(\mathcal{A})
\end{equation}
where $\mathcal{I}_{\mathbf{k}}$ refers to the indices of the selected core behaviors. Notably, the core behavior selection operation is non-differentiable, which impedes gradient propagation during backpropagation and can lead to training instability. To circumvent this issue, we employ the Straight-Through Estimator (STE)\cite{ste}  to ensure stable model training. Specifically, we generate a binary mask $\mathbf{M} \in \{0, 1\}^{L}$ where the top-$k$ indices of the importance score distribution $\mathcal{A}$ are set to 1. In the forward pass, a discrete mask is applied to the sequence to effectively select only the core behaviors for the subsequent feature interactions. This approach significantly reduces computational overhead, ensuring efficient online inference. In contrast, during the backward pass of offline training, to overcome the non-differentiable nature of this discrete selection, we treat the mask as a differentiable probabilistic approximation to ensure that model parameters are updated based on all behaviors in the user sequence:
\begin{equation}
\mathcal{S}_{core}  = \text{Gather}\left( \mathcal{S} \odot (\text{sg}[\mathbf{M} - \mathcal{A}] + \mathcal{A}), \mathcal{I}_{\mathbf{k}} \right)
\end{equation}
where $\text{sg}[\cdot]$ denotes the stop-gradient operation.

\subsection{Global Interest-Distribution Module}
While subjecting non-core behaviors to fine-grained interaction is prone to introducing noise, entirely disregarding them would lead to a significant loss of behavior information. To reconcile this trade-off, we propose modeling the user's holistic interest distribution as captured from the full behavior sequence. This coarse-grained representation serves to compensate for the information sacrificed by excluding these non-core behaviors from feature interaction.

Specifically, we first divide the range of similarity values from $0$ to $1$ equally into $n$ similarity range intervals, denoted as $c_1=[0, \frac{1}{n}]$, $c_2=(\frac{1}{n}, \frac{2}{n}]$, ..., and $c_n=(\frac{n-1}{n}, 1]$. Then, we calculate the number of behaviors with similarity scores in the user sequence as follows:
\begin{gather}
p_{j} = \# \left\{a_k | ( a_k \in c_j ) \right\}
\end{gather} 
where $p_{j}$ denotes the number of similarity scores between item $i$ and historical behaviors in the entire sequence within the range of $c_j$. $\#\left\{ \cdot \right\}$ represents the number of instances that satisfy a certain condition. If most of the behaviors in the sequence have a high similarity with the target item, it indicates that the user has a high interest in the target item. Using the above method, we can construct global user interest distributions based on the user sequence, denoted as $\mathcal{S}_{gid}=\frac{\sum_{1 \leq i \leq n} (\boldsymbol{p}_{i})}{n}$, where $\boldsymbol{p}_{i}$ is the embedding of $p_{i}$.

\subsection{Feature Interaction and Prediction Layer}
Following the extraction of core behaviors and the modeling of the user's interest distribution, we employ a standard Transformer architecture to perform the subsequent feature interactions and generate the final prediction. Specifically, we concatenate $k$ core behaviors of $\mathcal{S}_{core}$, the single interest distribution $\mathcal{S}_{gid}$, and the contextual features $\mathcal{F}$ to form the initial input $\mathbf{X}_{0} \in \mathbb{R}^{k+1+N_f}$ for the interaction network, where $N_f$ denotes the number of contextual features. The unified feature interaction is then defined as follows:
\begin{equation}
    \hat{\mathbf{X}}_{h} = \text{LN}(\text{MHA}(\mathbf{X}_{h-1}) + \mathbf{X}_{h-1})
\end{equation}
\begin{equation}
    \mathbf{X}_{h} = \text{LN}(\text{PFFN}(\hat{\mathbf{X}}_{h}) + \hat{\mathbf{X}}_{h})
\end{equation}
where $\text{LN}(\cdot)$ denotes Layer Normalization, MHA denotes the multi-head attention operation, $\text{PFFN}(\cdot)$ denotes the Per-token Feed-Forward Networks (PFFN)\cite{onetrans}. The final output $\hat{y}$ is obtained by applying an MLP to the output from the $H$-th layer, where $\hat y=MLP(\mathbf{X}_{H})$.

The optimization objective is the cross-entropy loss:
\begin{equation}
    \mathcal{L} = -\frac{1}{|\mathcal{D}|} \sum_{u,i \in \mathcal{D} }^{|\mathcal{D}|} \left( y \log(\hat{y}) + (1 - y) \log(1 - \hat{y}) \right)
\end{equation}
where $\mathcal{D}$ represents the training dataset.

\section{Experiments}
In this section, we conduct extensive experiments on public and industrial datasets to evaluate the effectiveness of CDNet.
\subsection{Experiment Settings}
\subsubsection{Dataset}
To validate the effectiveness of the proposed framework, we conducted experimental evaluations on a real-world large-scale industrial dataset as well as public benchmark datasets.

\textbf{Taobao}\footnote{\url{https://tianchi.aliyun.com/dataset/dataDetail?dataId=649}}:  It is based on user behavior logs from Taobao, one of the major e-commerce platforms in China, and contains 89 million records. We take the click behaviors for each user and sort
them according to time to construct the behavior sequence. We use the latest 600 clicked items as behavior sequential features. 

\textbf{Industrial}: The Industrial dataset is sourced from a real-world e-commerce platform, containing 7 billion user interactions and covering diverse product features as well as user behavior sequences. The user behavior sequence reaches a maximum length of 1,600. 

\subsubsection{Evaluation Metrics}
To evaluate prediction accuracy, we adopt three widely used industry metrics: AUC, GAUC\cite{GAUC}, and LogLoss. 

\subsubsection{Baselines}
We evaluate the proposed framework against mainstream feature interaction paradigms, which are categorized into :1) aggregate scheme with three sequence aggregation strategies, including \emph{Mean Pooling}, \emph{Taget-attention Pooling}, and \emph{Self-attention Pooling}, under DCN\cite{dcn}, AutoInt\cite{AutoInt}, Hiformer\cite{gui2023hiformerheterogeneousfeatureinteractions}, Wukong\cite{Wukong}, and RankMixer\cite{RankMixer}; 2)Joint interactions between sequential and contextual features scheme, including InterFormer~\cite{interformer} and OneTrans~\cite{onetrans}.

\subsubsection{Implementation Details}
We set $k=16$ for behavioral distillation and a similarity interval $n = 5$. The sequence length is set to 600 for the Taobao dataset and 256 for the industrial dataset in the overall performance comparison. %To further investigate the scalability of our model. % Experiment~\ref{exp:long_sequence} evaluates the impact of extended sequence lengths on the industrial dataset, with interactions varying across 256, 1,600.

\begin{table}[htbp] 
%\vspace{-0.3cm}
  \centering
  \caption{Overall performance comparison in CTR prediction on public and industrial datasets. "Improv." denotes the relative improvement of CDNet over the best baseline. The best baseline performance score is denoted in \underline{underline}.}
  \label{tab:overall}
  \resizebox{\columnwidth}{!}{% 
    \begin{tabular}{llcccccc} 
      \toprule
      \multicolumn{2}{c}{\textbf{Dataset}} & \multicolumn{3}{c}{\textbf{Taobao}} & \multicolumn{3}{c}{\textbf{Industrial}} \\
      \cmidrule(lr){3-5} \cmidrule(lr){6-8}
      \multicolumn{2}{c}{\textbf{Method}} & \textbf{AUC} & \textbf{GAUC} & \textbf{Logloss} & \textbf{AUC} & \textbf{GAUC} & \textbf{Logloss} \\
      \midrule
      \midrule
       \multicolumn{2}{l}{DCN+Mean Pool.} & 0.6299 & 0.6255 & 0.1962 & 0.7128 & 0.6121 & 0.0873 \\
     \multicolumn{2}{l}{DCN+Target-attention Pool.} & 0.6303 & 0.6261 & 0.1960 & 0.7135 & 0.6129 & 0.0869 \\
      \multicolumn{2}{l}{DCN+Self-attention Pool.} & 0.6315 & 0.6267 & 0.1957 & 0.7141 & 0.6137 & 0.0864 \\
       \midrule
      \multicolumn{2}{l}{AutoInt+Mean Pool.} & 0.6327 & 0.6271 & 0.1953 & 0.7147 & 0.6131 & 0.0869 \\
     \multicolumn{2}{l}{AutoInt+Target-attention Pool.} & 0.6331 & 0.6276 & 0.1951 & 0.7142 & 0.6132 & 0.0872 \\
      \multicolumn{2}{l}{AutoInt+Self-attention Pool.} & 0.6329 & 0.6269 & 0.1953 & 0.7158 & 0.6149 & 0.0860 \\
       \midrule
      \multicolumn{2}{l}{Hiformer+Mean Pool.} & 0.6347 & 0.6285 & 0.1952 & 0.7147 & 0.6147 & 0.0853 \\
     \multicolumn{2}{l}{Hiformer+Target-attention Pool.} & 0.6352 & 0.6293 & 0.1948 & 0.7148 & 0.6155 & 0.0852 \\
      \multicolumn{2}{l}{Hiformer+Self-attention Pool.} & 0.6349 & 0.6288 & 0.1950 & 0.7145 & 0.6152 & 0.0853 \\
       \midrule
    \multicolumn{2}{l}{Wukong+Mean Pool.} & 0.6341 & 0.6301 & 0.1955 & 0.7213 & 0.6182 & 0.0849 \\
     \multicolumn{2}{l}{Wukong+Target-attention Pool.} & 0.6349 & 0.6309 & 0.1950 & 0.7201 & 0.6176 & 0.0849 \\
      \multicolumn{2}{l}{Wukong+Self-attention Pool.} & 0.6350 & 0.6304 & 0.1949 & 0.7202 & 0.6180 & 0.0850 \\
       \midrule
    \multicolumn{2}{l}{RankMixer+Mean Pool.} & 0.6336 & 0.6297 & 0.1955 & 0.7207 & 0.6175 & 0.0868 \\
    \multicolumn{2}{l}{RankMixer+Target-attention Pool.} & 0.6341 & 0.6301 & 0.1958 & 0.7215 & 0.6182 & 0.0865 \\
    \multicolumn{2}{l}{RankMixer+Self-attention Pool.} & 0.6340 & 0.6298 & 0.1956 & 0.7221 & 0.6190 & 0.0860 \\
    \midrule
    \multicolumn{2}{l}{InterFormer} & 0.6336 & 0.6281 & 0.1959 & 0.7198 & 0.6174 & 0.0869 \\
     \multicolumn{2}{l}{OneTrans} & \underline{0.6351} & \underline{0.6319} & \underline{0.1947} & \underline{0.7239} & \underline{0.6214} & \underline{0.0843} \\
     % \multicolumn{2}{l}{Hyformer} & - & - & - & - & - & - \\
      \multicolumn{2}{l}{\textbf{CDNet}} & \textbf{0.6388} & \textbf{0.6340} & \textbf{0.1943} & \textbf{0.7248} & \textbf{0.6230} & \textbf{0.0836} \\
      \midrule
      \multicolumn{2}{l}{\textbf{Improv.}} & \textit{ +0.58\% } & \textit{+0.33\%} &\textit{ -0.21\% }& \textit{+0.12\%} &\textit{+0.26\%} & \textit{-0.83\%} \\
      \bottomrule
    \end{tabular}%
  }
\end{table}

\subsection{Overall Performance}
Table \ref{tab:overall} presents the overall predictive performance of all methods on both industrial and public datasets, alongside the relative improvement of our method over the best-performing baseline. The results across both datasets robustly demonstrate the superiority of our proposed method, CDNet. Specifically, OneTrans consistently outperforms methods that adhere to the traditional two-stage paradigm (i.e., sequence aggregation followed by feature interaction). This underscores the necessity of modeling fine-grained interactions between user sequences and contextual features. However, the full sequence expansion strategy employed by OneTrans inevitably introduces significant noise from irrelevant behaviors. This, in turn, can obscure the crucial interaction signals from the most relevant behaviors. This inherent limitation explains why CDNet, with its core-behavior selection mechanism, achieves superior performance.

\subsection{Ablation Experiments}
To ablate the contributions of the components in CDNet, we designed two variants: 1) CDNet-RCore, which omits the core behaviors, relying solely on the interest distribution. As shown in Table \ref{tab:ablation}, its performance drops substantially, validating our hypothesis that modeling fine-grained interactions with core behaviors is critical for predictive accuracy; 2) CDNet-RGid, which removes the interest distribution, considering only the core behaviors. Although the resulting performance degradation is less severe than that of CDNet-RC, the decline is still significant. This validates the importance of the interest distribution as a vital compensatory mechanism for information lost during the core behavior selection process. Taken together, these findings demonstrate that fine-grained and coarse-grained components are highly complementary, with both being indispensable for achieving CDNet's optimal performance.

\subsection{Hyper-parameters Analysis}
\label{exp:long_sequence}
We investigate the impact of key hyperparameters on model performance: 1) The Impact of Selection Ratio of Core Behaviors ($k/L$): We examine the sensitivity of the core behavior selection ratio $k/L$ under different sequence scales, specifically $L \in \{256, 1600\}$. Utilizing the full sequence ($k/L = 1$) results in sub-optimal performance compared to the peak selection ratio. This confirms that raw, long-range user histories are inherently plagued by substantial irrelevant noise, which can distract the model and compromise its effectiveness. These findings validate the necessity of our selection mechanism in selecting core behaviors; 2) Impact of f Similarity Intervals $n$: Through experiments with various similarity intervals, we observe that smaller $n$ consistently yield superior performance. This suggests that increased granularity of the interest distribution enables the model to capture user interests with higher precision, effectively characterizing users' complex behavioral patterns.

\begin{table}[tbp]
  \centering
  \caption{Ablation study of CDNet.}
  \vspace{-0.3cm}
  \label{tab:ablation}
  \small
 % \resizebox{0.8\columnwidth}{!}{% 
  \begin{tabular}{lccc}
    \toprule
    \multicolumn{1}{l}{\textbf{Variants}} & \textbf{AUC} & \textbf{GAUC} & \textbf{Logloss} \\
    \midrule
      \textbf{CDNet} & \textbf{0.7248} &\textbf{0.6230} & \textbf{0.0836} \\
       CDNet-RCore & 0.7230 & 0.6203 & 0.0850 \\
       CDNet-RGid & 0.7240 & 0.6214 & 0.0842 \\
    \bottomrule
  \end{tabular}
 % }
\end{table}

\begin{figure}[t]
\includegraphics[width=0.9\columnwidth]{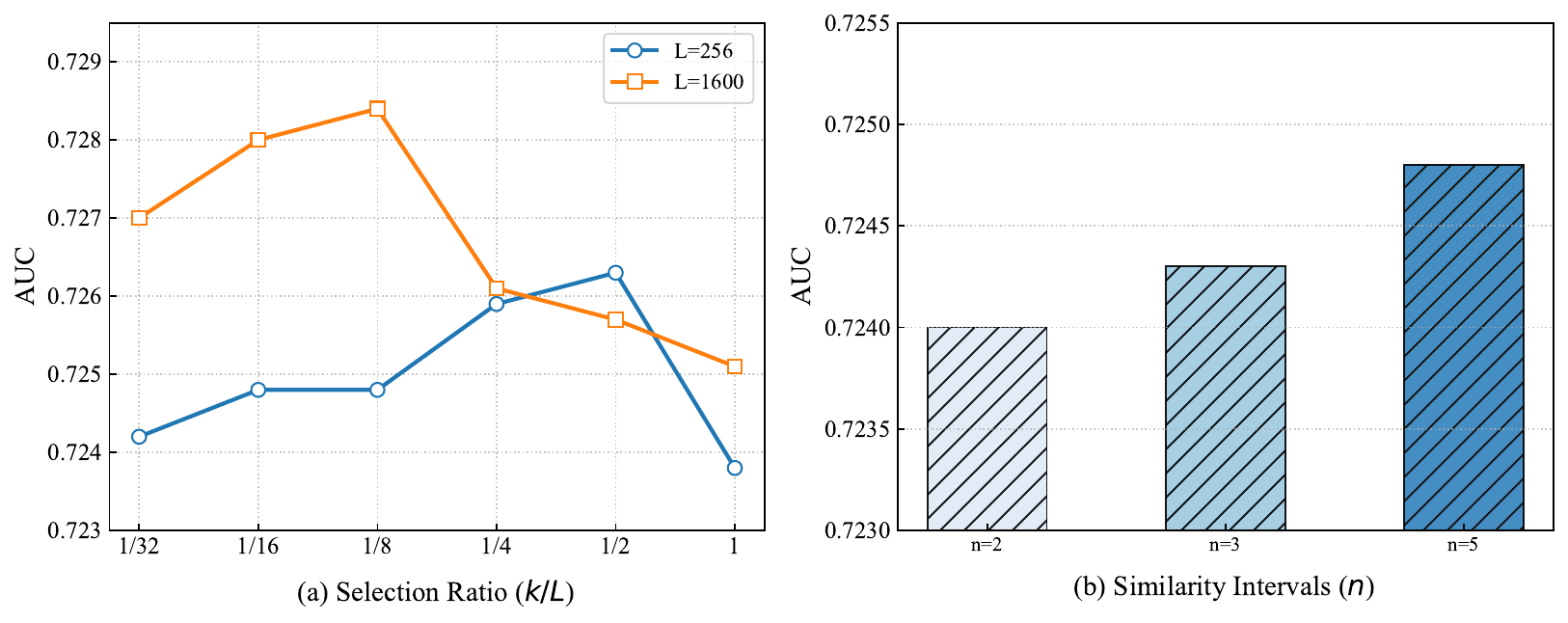}
\centering
\caption{Hyper-parameters analysis of CDNet}
\label{fig:intro_heat}
\vspace{-0.5cm}
\end{figure}

\section{Online Experiments}
To conclusively validate the real-world efficacy of our proposed method, we conduct a 10-day online A/B test on a large-scale e-commerce platform. Compared to the incumbent production model, CDNet achieved a 2.24\% relative lift in CTR with the user behavior sequence length of 1,600 and 100 core behaviors, while incurring zero additional inference latency. This demonstrates CDNet's practical efficiency for real-time large-scale recommendation systems.

\textbf{Complexity Analysis}
The computational complexity of the feature interaction layer in CDNet is $\mathcal{O}((k + 1 + N_f)^2)$, where $k$, $1$, and $N_f$ represent the number of core behaviors, global interest distribution, and contextual features, respectively. 
Compared to state-of-the-art interaction methods like OneTrans, which incur a complexity of $\mathcal{O}((L + N_f)^2)$, CDNet is significantly more efficient. 

\section{Conclusion}
In this paper, we introduce the Core-Behaviors and Distributional-Compensation Dual-View Interaction Network (CDNet), a novel framework designed to model the intricate interactions between user behavior sequences and non-sequential contextual features. CDNet operates from two complementary perspectives: 1) A fine-grained view that explicitly models interactions between a select set of core historical behaviors and the contextual features, thereby mitigating noise from irrelevant behaviors for the candidate item. 2) A coarse-grained view that models the user's global interest distribution, compensating for the contextual information sacrificed during the fine-grained core behavior selection process. By synergizing the above two behavior views, CDNet effectively captures feature interactions with high precision while remaining computationally efficient and robust to behavior noise. Extensive experiments on both public and industrial datasets robustly demonstrate the superiority of CDNet and validate the synergistic contribution of its dual-view design for bridging sequential and contextual features.

\bibliographystyle{ACM-Reference-Format}
\bibliography{main}
\end{document}